\documentclass[prl,aps,superscriptaddress,noshowpacs,twocolumn,reprint,noshowkeys,amssymb,amsmath,amsfonts]{revtex4-1}

\usepackage{graphicx}
\usepackage{theorem}
\usepackage{dcolumn}
\usepackage{longtable}
\usepackage{bm}
\usepackage{accents}

\newcommand{\SSS}{\scriptscriptstyle}
\newcommand{\DS}{\displaystyle}

\newcommand{\Ex}{\hat{\mathbf{e}}_{x}}

\newcommand{\Ez}{\hat{\mathbf{e}}_{z}}

\newcommand{\kB}{k_{\SSS\text{B}}}
\newcommand{\muB}{{\mu_{\SSS\text{B}}}}

\newcommand{\specialcell}[2][c]{%
	\begin{tabular}[#1]{@{}c@{}}#2\end{tabular}}

\makeatother

\begin{document}

\title{Quantum Electronic Structure at the Interface of Solid Neon and Superfluid Helium}

\author{Dafei Jin}\email{djin@anl.gov}
\affiliation{Center for Nanoscale Materials, Argonne National Laboratory, Argonne, Illinois 60439, USA}

\date{\today}

\begin{abstract}
We predict a new quantum electronic structure at the interface between two condensed phases of noble-gas elements: solid neon and superfluid helium. An excess electron injected onto this interface self-confines its wavefunction into a nanometric dome structure whose size varies with pressure. A collection of such electrons can form a classical Wigner crystal visualizable by mid-infrared photons. The ultralong spin-coherence time allows the electrons in this system to serve as perfect quantum bits. They can be deterministically arranged on-chip at a spacing of several microns. Their spin states can be controlled and readout by single-electron devices. This unique system offers an appealing new architecture for scalable quantum information processing.
\end{abstract}

\maketitle
\pretolerance=8000 

Noble-gas elements helium (He) and neon (Ne) are the two most stable chemical elements in the universe. When an excess electron (e$^{-}$) approaches helium or neon (He/Ne) atomic cores, it experiences a strong repulsive force due to the Pauli exclusion from the occupied orbital electrons. To enter a condensed liquid or solid He/Ne at low temperature, the excess electron must overcome a potential barrier of the order of 1~eV~\cite{sommer1964liquid,springett1968stability,cohen1969electron,miyakawa1969stability,loveland1972experimental}. Inside bulk He/Ne, the ground-state electron pushes away all the surrounding atoms and opens up a nanometric cavity, known as an electron bubble~\cite{maris2003properties,jin2010electrons}. On a He/Ne-to-vacuum surface, the ground-state electron is weakly attracted by its image charge and forms a dimple structure~\cite{cole1969image}. The dimple radius is over 500~nm on liquid He surface~\cite{grimes1979evidence} and much larger on solid Ne surface~\cite{kajita1983stability}. A collection of electrons on these surfaces can form a classical two-dimensional electron gas (2DEG)~\cite{grimes1979evidence,kajita1983stability}.


In the past decades, there has been considerable interest of using the long-coherence orbital and spin states of the electrons on liquid He surface to engineer quantum bits (qubits)~\cite{platzman1999quantum,papageorgiou2005counting,lyon2006spin,sabouret2008signal,schuster2010proposal,yang2016coupling}. In particular, since $^4$He (excluding $^3$He) has zero atomic spin and constitutes an ultraclean superfluid below 1~K~\cite{leggett2006quantum}, electrons in this environment possess a spin-coherence time over 100~s~\cite{lyon2006spin,yang2016coupling}. Likewise, $^{20}$Ne and $^{22}$Ne (excluding $^{21}$Ne) also have zero atomic spin, and can form ultrapure solids below 23~K~\cite{henshaw1958atomic,loveland1972experimental}. They can potentially serve as a new solid-state storage matrix for qubits. However, by far it is still challenging to precisely position every electron, control and readout its spin states on either of these surfaces. The large dimple radius can result in overlapped electron wavefunctions in nanoscale devices~\cite{yang2016coupling,koolstra2019spectroscopy}.

In this paper, we reveal a new quantum electronic structure at the interface between solid Ne and superfluid He. We show that each electron is self-confined into a dome structure, whose flat side attaches to the solid Ne and curved side dips into the superfluid He. Employing a bosonic density functional theory (DFT) which can accurately reproduce essential properties of superfluid He and electron bubble~\cite{stringari1987surface,jin2010vortex,mateo2011excited}, we calculate the ground state, excited states, and optical transitions. The dome diameter decreases with increasing pressure from 7~nm at 1~bar to 3~nm at 25~bar. The optical transition wavelengths are in the mid-infrared (mid-IR) regime. When a number of electrons are deposited at the interface and bounded by a hard-wall potential, they can form a classical Wigner crystal. The unusual feature is the possibility to visualize such Wigner crystal with mid-IR photons. Finally, we propose to employ this structure for quantum information processing. Each electron can be trapped by a hole-array electrode with 10~$\mu$m spacing. It allows for developing on-chip electron paramagnetic resonance (EPR) devices~\cite{pla2012single}. Superconductor nanowires (SNW) and single electron transistors (SETs)~\cite{papageorgiou2005counting,koolstra2019spectroscopy,pla2012single} can be used to control and readout the spin states.

\begin{figure}[hbt]
\centerline{\includegraphics[scale=0.8]{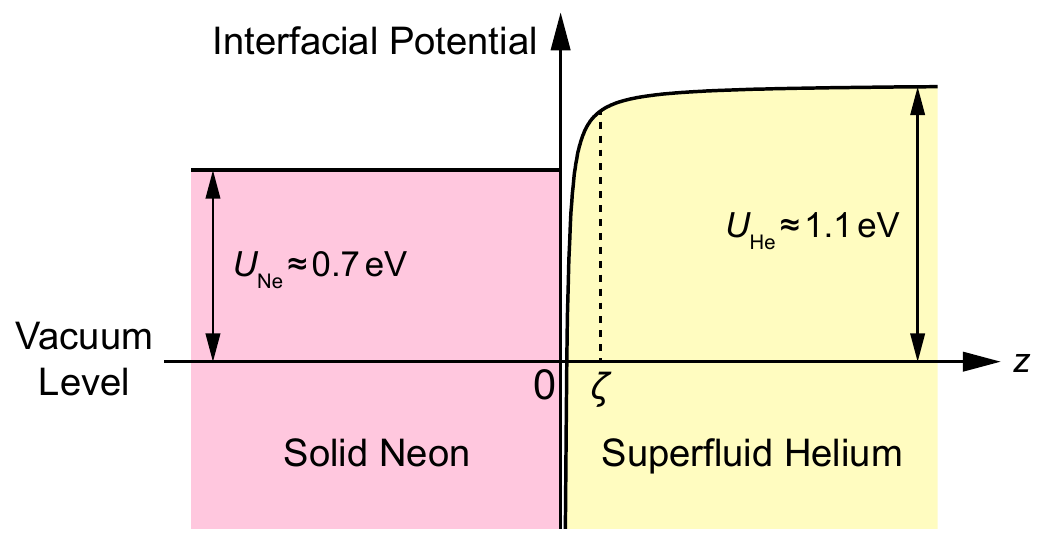}}
\caption{Interfacial potential for an electron sandwiched between flat superfluid He and solid Ne at zero pressure.}
\end{figure}

\begin{figure*}[hbt]
\centerline{\includegraphics[scale=0.8]{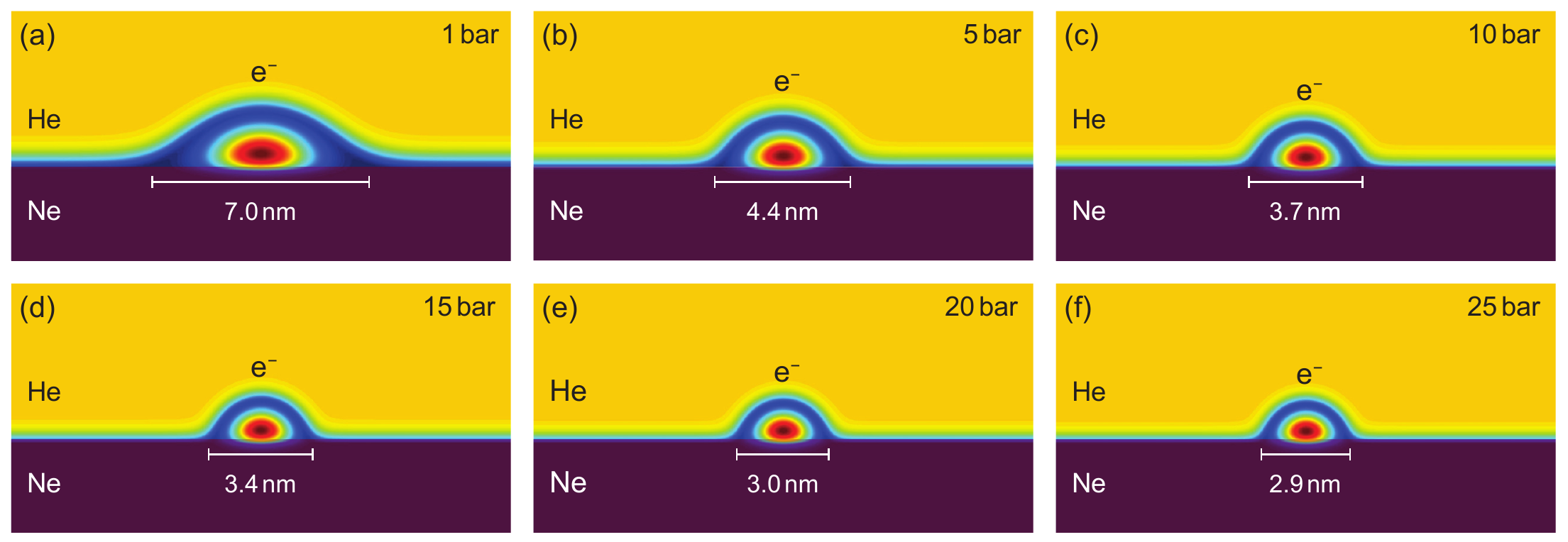}}
\caption{Calculated ground-state electron and helium density profiles of the single-electron dome structure at the interface between superfluid He and solid Ne under 1 to 25~bar pressure.}
\end{figure*}

Figure~1 gives the DFT calculated interfacial potential undergone by an electron sandwiched between flat solid Ne and superfluid He. It consists of three contributions: the Pauli-exclusion potential barrier from Ne and He, respectively, and the image-charge attractive potential from Ne. It can be approximated as
\begin{equation}
U(z) =
\begin{cases}
U_{\text{Ne}}, &\  z<0, \\
\DS -\frac{\epsilon_{\text{Ne}}-1}{\epsilon_{\text{Ne}}+1}\frac{e^2}{4z} + U_{\text{He}} \tanh^2\left(\frac{z}{\zeta}\right), &\  z>0.
\end{cases}
\end{equation}
Here, $U_{\text{He}}\approx 1.1$~eV and $U_{\text{Ne}}\approx0.7$~eV are the bulk potential barriers of superfluid He and solid Ne to the electron, and $\epsilon_{\text{Ne}} = 1.244$ is the dielectric constant of solid Ne~\cite{cole1969image}. Since the polarizability of He is much smaller compared with Ne, we completely ignore the polarization effect from superfluid He~\cite{maris2003properties,jin2010vortex}, and simply take its dielectric constant as that of vacuum, $\epsilon_{\text{He}} = 1.056\approx 1$. Among the above parameters, only $U_{\text{He}}$ varies appreciably with pressure $p$ and is counted in our calculation. Unlike the previously studied Ne-vacuum interface~\cite{kajita1984new}, here the strong repulsion from He overrides the weak and long tail of polarization potential from Ne. Therefore, an attractive potential only exists in a narrow region $0<z<\zeta$ of the order of healing length of superfluid He, which is only about 1~{\AA}~\cite{leggett2006quantum,jin2010vortex}.

\begin{table*}[htb]
\caption{Calculated electronic and optical properties of single-electron dome structure under various pressures.}
\begin{center}
\begin{ruledtabular}
\begin{tabular}{c|ccccccc}
Pressure (bar) & \specialcell{Helium\\ Barrier (eV)} & \specialcell{Dome\\ Diameter (nm)} & \specialcell{Dome\\ Height (nm)} & \specialcell{$l_z=0, n_r=1$ \\ Electron Energy (eV)} & \specialcell{$l_z=1, n_r=1$ \\ Electron Energy (eV)} & \specialcell{Transition\\ Wavelength ($\mu$m)}\\
\hline
0 & 1.05 & $\infty$ & 0 & 0 & 0 & $\infty$ \\
1 & 1.07  & 6.96 & 2.15 & 0.105 & 0.156 & 24.3\\
5 & 1.10  & 4.38 & 1.84 & 0.154 & 0.245 & 13.6 \\
10 & 1.15 & 3.66 & 1.67 & 0.188 & 0.304 & 10.7\\
15 & 1.18 & 3.36 & 1.56 & 0.212 & 0.347 & 9.18 \\
20 & 1.21 & 3.02 & 1.48 & 0.232 & 0.381 & 8.32 \\
25 & 1.24 & 2.90 & 1.43 & 0.249 & 0.411 & 7.65 \\
\end{tabular}
\end{ruledtabular}
\end{center}
\end{table*}

Without an applied pressure, the electron wavefunction between He and Ne is unbounded in the $xy$-plane. A small positive pressure immediately confines the electron. The rigid solid Ne and the soft superfluid He wrap the single-electron wavefunction into a dome. Figure~2 shows the DFT calculated dome structure, distinguished from the dimple structure or bubble structure on the surface or in the bulk of superfluid He. The system preserves the circular symmetry in the $xy$-plane but no apparent symmetries along the $z$-axis. The orbital angular momentum $l_z=0,\pm1,\pm2,\dots$ is a good quantum number. $|l_z|$ corresponds to the number of nodal surfaces in the azimuthal direction. For every given $l_z$, there is the radial quantum number $n_r=1, 2, 3, \cdots$. $n_r-1$ corresponds to the number of nodal surfaces in the radial direction. As displayed in Fig.~2, an increasing pressure from 1 to 25~bar (below the liquid-solid phase transition of $^4$He) shrinks the dome diameter from 7~nm to 2.9~nm and dome height from 2.15~nm to 1.43~nm.

\begin{figure}[hbt]
	\centerline{\includegraphics[scale=0.8]{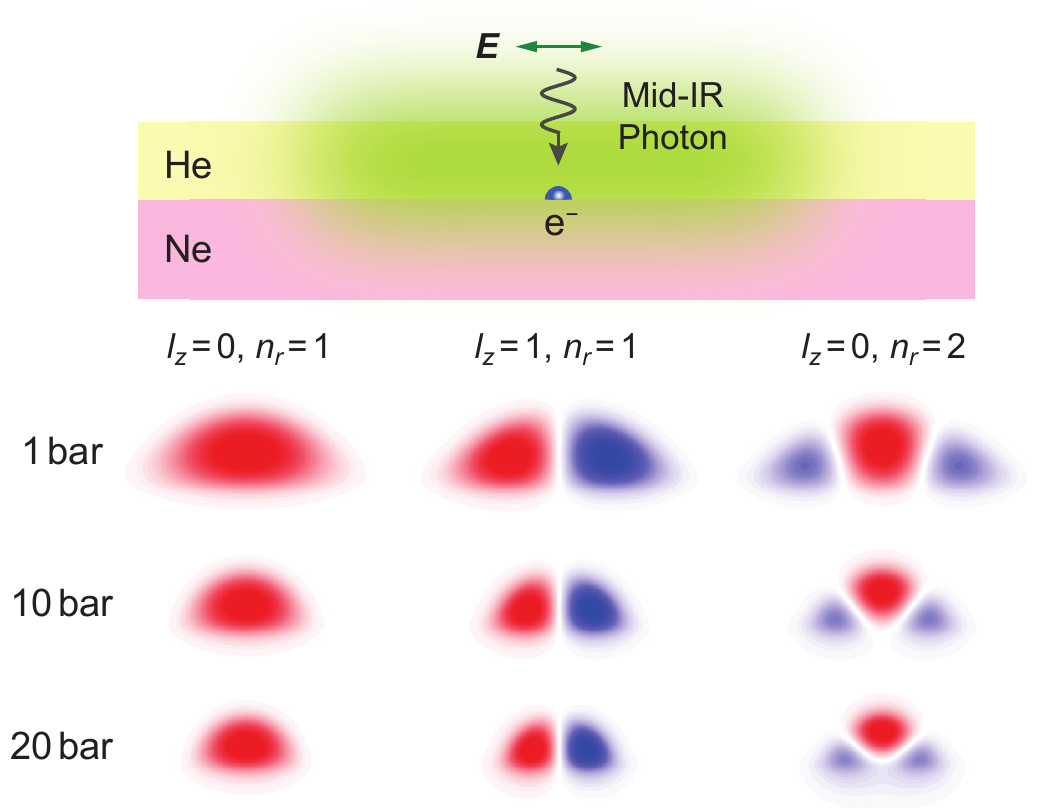}}
	\caption{Schematics of incident mid-infrared photons and calculated eigen-wavefunctions of a single electron in the dome structure under 1, 10 and 20~bar pressure.}
\end{figure}

\begin{figure}[hbt]
	\centerline{\includegraphics[scale=0.8]{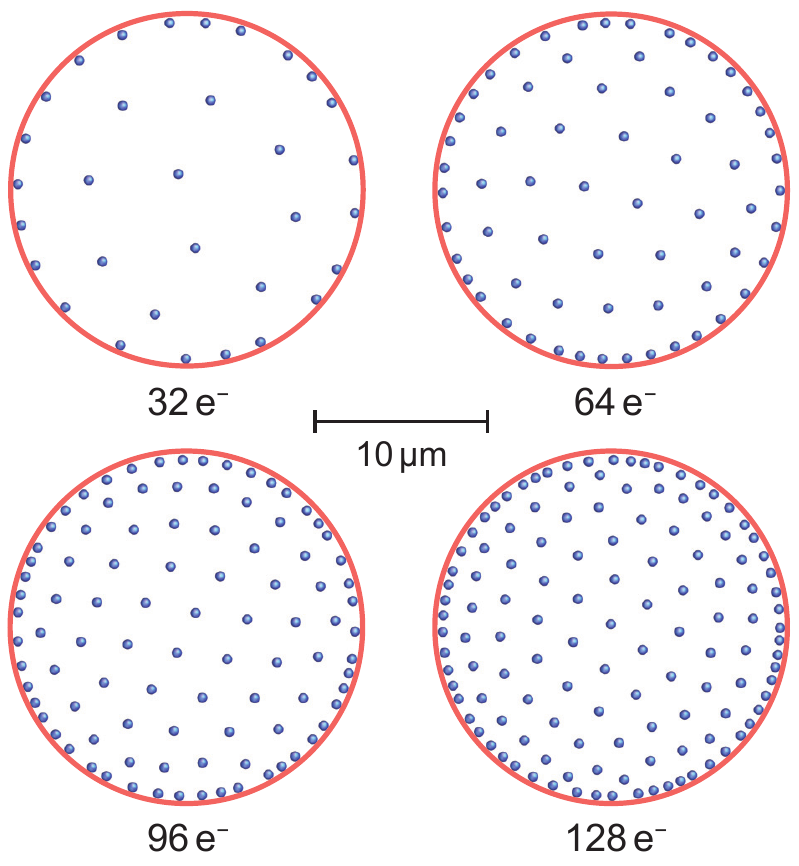}}
	\caption{Formation of a classical 2D Wigner crystal confined by a 10~$\mu$m radius hard-wall potential at the interface of solid Ne and superfluid He.}
\end{figure}

When linearly (or circularly) polarized photons are normally incident onto the electron dome, they can pump the electron from the ground state to excited states~\cite{grimes1992infrared,maris2003properties,jin2010electrons}. According to the Franck-Condon principle, here the optical transitions happen in the femtosecond timescale, much faster than the deformation of superfluid He in the picosecond timescale~\cite{jin2010electrons}. During the transition, the electron sees a frozen Hamiltonian corresponding to the ground-state dome cavity. Among many possible transitions, the 1-photon transition: $l_z=0$, $n_r=1$ to $l_z=\pm1$, $n_r=1$, is the dominant selection-rule allowed transition and has the largest transition dipole moment. The 2-photon transition: $l_z=0$, $n_r=1$ to $l_z=0$, $n_r=2$, is the selection-rule forbidden transition; the final state maintains the circular symmetry and contains a nodal surface in the radial direction. Figure~3 shows the calculated electron wavefunctions for several eigenstates under representative pressures: 1~bar, 10~bar and 20~bar. The electron eigen-energies and 1-photon optical transition wavelengths are calculated and listed in Table~I. The transition wavelengths are all in the mid-IR spectral range, varying with pressure. In experiment, one can use a broadband thermal source or quantum cascade laser to excite the electron. Especially at around 10~bar, the wavelength 10.7~$\mu$m is near the emission wavelength of the strong CO$_2$ laser. To our knowledge, no other (quasi)2D electronic systems~\cite{platzman1999quantum,koppens2006driven} exhibit similar lateral quantum confinement on resonance with visible or infrared photons.

Our system offers a new platform to realize classical 2D Wigner crystal~\cite{grimes1979evidence,rousseau2009addition,andrei1988observation,zhu2010observation}. Although each individual electron wavefunction is extremely quantum confined, its spatial motion as a whole is classical point-particle alike, carrying a hydrodynamic effective mass of a few hundreds of He atoms~\cite{shikin1977mobility}. The relatively low electron density $n_e$ makes the quantum degeneracy temperature about $10^{-5}$~K only. Even at an ambient temperature $T$ of 10~mK, as that provided by a dilution refrigerator, classical statistics still dominates over quantum statistics. The $\Gamma$ parameter quantifying the disorder to order transition of a classical Wigner crystal is
\begin{equation}
\Gamma = \frac{e^2\sqrt{\pi n_e}}{\kB T}. \label{Eqn:Wigner}
\end{equation}
Previous experiment on liquid He surface shows $\Gamma\approx 137$ in agreement with the Monte Carlo simulations~\cite{grimes1979evidence,rousseau2009addition}.

\begin{figure*}
	\centerline{\includegraphics[scale=0.8]{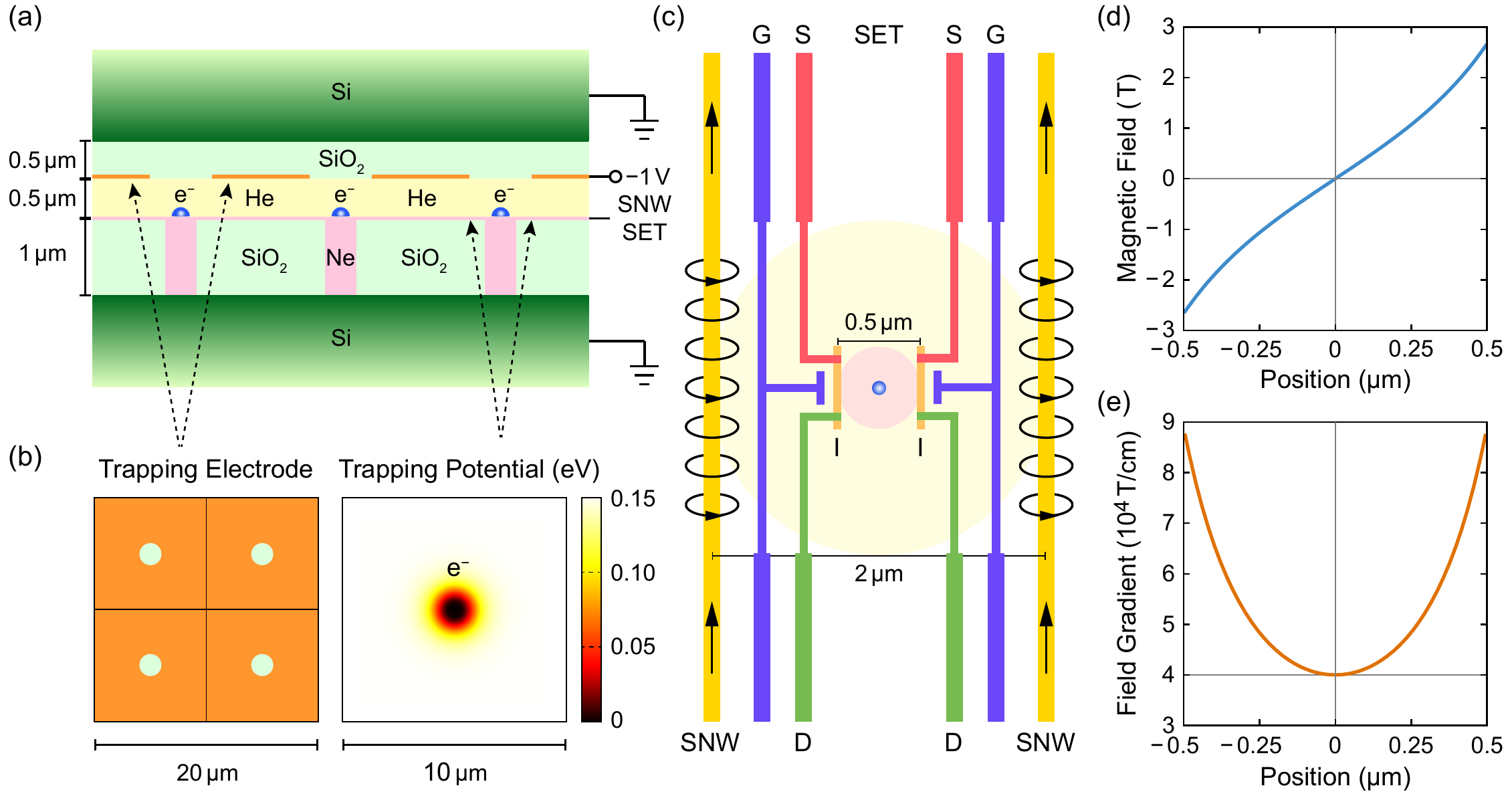}}
	\caption{Schemes for trapping single electrons into an ordered array and performing single electron qubit manipulation. (a) Design of an electrostatic trap. (b) Simulation of the trapping potential. (c) Design of the superconductor nanowires (SNWs) and single-electron transistors (SETs) for the control and readout of electron spin. The electrodes are labeled as: gate (G), drain (D), source (S), and island (I). (d) Calculated magnetic field and its gradient by the SNWs. }
\end{figure*}

Here we calculate the Wigner crystal phase at the interface of solid Ne and superfluid He. We adopt a classical Monte Carlo simulation \cite{rousseau2009addition} for $N$ Newtonian electrons exposed to the Coulomb interaction, friction force, and a random force $f(t)$. The latter two obey the classical fluctuation-dissipation relation $\langle f(t)f(t') \rangle = 2 \eta\kB T \delta(t-t')$, in which $\eta=\gamma m_*$ is the viscous coefficient; $\gamma$ is the damping rate and $m_*$ is the effective mass. We choose the ambient temperature $T=100$~mK and a proper $\eta$ that quickly damps the system towards equilibrium. Eq.~(\ref{Eqn:Wigner}) indicates a critical electron number density $2.1\times10^7$~cm$^{-2}$. We set up a hard-wall potential to prevent electrons from escaping outside a circle of 10~$\mu$m radius. This gives a critical electron number about $67$ inside. We compare the results with different electrons in the circle. Figure~4 does suggest that above about $64$, a hexagonal order of the electrons begins to develop. The unique feature of this Wigner crystal, compared with all existing ones, is the possibilty to be visualized by mid-infrared absorption imaging~\cite{celebrano2011single}. To date, this is the only Wigner crystal that has on-resonance response to optical photons and can possibly be directly visualized. Moreover, since the 2DEG in this system does not rely on the polarization potential or the breakdown electron density on liquid He surface, experimental realization can be done at much higher temperature such as 1~K, with much higher electron density $n_e\gtrsim10^{10}$~cm$^{-2}$~\cite{kajita1984new}.

Lastly, we consider using this unique system as a new on-chip quantum information platform. The nanometric electron dome structure at the He-Ne interface enables single-electron manipulation down to 10~nm length scale, much finer than the conventional dimple structure at the He-vacuum interface. We can easily trap the electrons into a spatially ordered array. Since the barriers from superfluid He and solid Ne have provided natural trapping in $z$, there is no need to make a radiofrequency or magnetic trap~\cite{major2006charged}. An in-plane electrostatic trap is sufficient. Figure~4(a) gives a design. A heavily doped silicon (Si) with a 1~$\mu$m thick silicon dioxide (SiO$_2$) on top can be used as a substrate. An array of holes with about 500~nm diameter is etched into the SiO$_2$ and filled with Ne. Above Ne, there is an about 500~nm thick superfluid He. Electronic control and readout devices can be fabricated on the SiO$_2$, flushing with the He-Ne interface. The height of Ne and He can be controlled by capacitance measurement. Above He, there is another Si/SiO$_2$ substrate placed up-side-down. An 100~nm thick gold (Au) electrode with 1~$\mu$m diameter holes are patterned on the SiO$_2$. By grounding the top and bottom Si electrodes and applying a small negative voltage about $-1$~V on the Au electrode, the center of each hole has an electron potential minimum at the interface. Our simulation in Fig.~4(b) suggests that a potential depth about 0.15~eV (namely, 1741~K) can be achieved. Compared with the typical experimental temperature below 1~K, this trapping is extremely strong and the electron will stay strictly at the center.

On the SiO$_2$ of the lower substrate, we can fabricate niobium (Nb) based SNWs and aluminum (Al) and aluminum oxide (Al$_2$O$_3$) based SETs to control and readout single-electron spins. A uniform magnetic field $B_0\Ez$ along $z$ from a superconductor coil can provide the Zeeman splitting $s_z=\pm\frac{1}{2}$ for all the electrons. Then one of the SNWs can be used to apply microwave pulses to bring a local electron onto an arbitrary spin state on the Bloch sphere following the pulsed EPR technique~\cite{pla2012single,wertz2012electron}.

To readout the electron spin, the high mobility of electrons in superfluid He enables a Stern-Glach setup. As shown in Fig.~5, a dc current $I$ can be applied to the two SNWs symmetrically at $x=\pm a$ and generate a space-varying field $B_1(x)\Ez=(2\pi)^{-1}\mu_0I[(a-x)^{-1}-(a+x)^{-1}]\Ez$. The electron experiences a force in $x$, $F(x)\Ex = \pm \muB \partial_x B_1(x)\Ex$, where the sign is determined by the spin.
As shown in Fig.~4(d-e), the gradient magnetic field when $I=1$~mA and $a=1$~$\mu$m can be as high as $4\times10^4$~T~cm$^{-1}$. This generates a force about $F=\pm 3.71\times 10^{-16}$~erg~cm$^{-1}$ between $x=\pm200$~nm. At below 50~mK, the electron bubble mobility in bulk liquid He approaches $\mu_{e} \sim 10^7$~cm$^2$~V$^{-1}$~s$^{-1}$~\cite{donnelly1998observed,yang2016coupling}, from which we can deduce a drift velocity under the magnetic force $v\sim 2.3\times 10^{3}$~cm~s$^{-1}$. No elementary excitations can be generated at this velocity since it is below the Landau critical velocity and vortex nucleation velocity~\cite{leggett2006quantum}. But it is so fast to only need 10~ns for the electron to travel about 230~nm distance to the superconductor island as shown in Fig.~4(c). 10~ns can be considered as the typical readout time, which is much faster than the spin decoherence time $T_1, T_2\gtrsim100$~s in this system~\cite{lyon2006spin}.

To conclude, we have demonstrated a new quantum electronic structure at the interface of superfluid He and solid Ne. Single electrons form a dome structure of only several nanometer size. They can absorb mid-IR photons of tunable wavelength with pressure. Below 1~K temperature, they can form a classical Wigner crystal, which can be visualized by mid-IR photons. The ultralong spin-coherence time in this system permits the development of electron-spin qubits. We propose the schemes to trap individual electrons into an ordered array and use on-chip single-electron devices to control and readout their spin states. This system can be of both fundamental interest and practical application in quantum information science.

\begin{acknowledgements}
This work was performed at the Center for Nanoscale Materials, a U.S. Department of Energy Office of Science User Facility, and supported by the U.S. Department of Energy, Office of Science, under Contract No. DE-AC02-06CH11357. The author thanks Humphrey Maris, Lloyd Engel, and Daniel Lopez for invaluable advice.
\end{acknowledgements}

\bibliography{ElectronAtHeNeInterfaceRef}

\end{document}